\documentclass[aps,pra,twocolumn,amsmath,amssymb,floatfix]{revtex4}

\usepackage{amsmath}
\usepackage{graphicx}
\usepackage{epsfig}
\usepackage[utf8]{inputenc}
\usepackage{bm}
\usepackage{bbm}
\usepackage{hyperref}
 \hypersetup{colorlinks=true}
 \usepackage{ulem}

\def\H1{\widehat{H}_1}

\newcommand{\pd}{\partial}
\newcommand\vvdots{\vphantom{\int^0}\smash[t]{\vdots}}
\newcommand{\pdag}{{\phantom{\dagger}}}
\usepackage{color}

\begin{document}

\title[]{Control over few photon pulses 
by a time-periodic modulation of the photon-emitter coupling}

\author{Mikhail Pletyukhov$^{1}$, Kim G. L. Pedersen$^{1}$, Vladimir 
Gritsev$^{2}$ }

\affiliation{
$^1$Institute for Theory of Statistical Physics and JARA -- Fundamentals of 
Future Information Technology, 
RWTH Aachen University, 52056 Aachen, Germany \\
$^2$Institute for Theoretical Physics, Universiteit van Amsterdam, Science Park 
904,
Postbus 94485, 1098 XH Amsterdam, The Netherlands
}
\begin{abstract}
We develop a Floquet scattering formalism for the description
of quasistationary states of microwave photons in a one-dimensional
waveguide interacting
with a nonlinear cavity by means of a periodically modulated coupling.
This model is inspired by the recent progress in engineering of tunable 
coupling schemes with superconducting qubits. We argue that our model
can realize the quantum analogue of an optical chopper.
We find strong 
periodic modulations of the transmission and reflection envelopes in the 
scattered few-photon pulses, including photon compression and blockade, as well 
as dramatic changes in statistics. Our theoretical analysis 
allows us to explain these non-trivial phenomena as arising from non-adiabatic 
memory effects. 
\end{abstract}

\maketitle

\section{Introduction.}

Periodically driven quantum systems -- or Floquet quantum systems as they are often called --  may behave markedly different than their equilibrium counterparts, and it has been shown time and time again that this difference in behavior serve a whole range of potential applications. 

 

In many-body quantum physics intensive research has recognized that the periodic driving of quantum many-body system could create new, synthetic phases of matter not accessible in equilibrium systems. This intuition, motivated by the classical example of the Kapitza pendulum ~\cite{Kapitza}, has been explored and confirmed in several contexts. In particular, some proposals predict the formation of topological phases~\cite{F6,kitagawa_11,KBRD} and artificial gauge systems \cite{Gold_Dali,F7,F16}, as well as localized non-thermal states in isolated many-body systems~\cite{ponte_15,bukov_16,foster_15,DP,bukov_15}.

In quantum information protocols proposals for dynamical decoupling schemes~\cite{VL,Ban,VKL,Z,KLV} and their refinements~\cite{KL, U,KV} use periodic sequences of fast and strong symmetrizing pulses to reduce the parts of the system-bath interaction Hamiltonian which are sources of decoherence. Additionally, Floquet systems also naturally appear in digital quantum computation schemes~\cite{F15}.

In quantum transport various Floquet-driven quantum tunneling problems~\cite{GH} are in the heart of physics described by an effective two-level systems, quantum wells and quantum open systems.





In this paper we seek to combine the possibilities offered by periodically driven quantum systems with the experimental flexibility available in quantum photonics, as e.g. realized in quantum optics or microwave quantum electro-dynamics.  
At this point it is important to stress that we do not simply talk about the time-dependence of e.g. a classical laser field where the time-dependence always trivially can be gauged away; instead we refer to quantum photonics systems where the time-dependence manifest itself directly in the steady-state observables, i.e. such that they themselves become time-dependent.

Specifically we are interested in the long time behavior of the observables which can be captured by a suitably formulated version of scattering theory.





In classical optics the most common periodically driven instruments are optical choppers and shutters~\cite{Choppers}, famous, perhaps, for their application in the first  non-astronomical speed of light measurements by Hippolyte Fizeau in 1849~\cite{Fizeau}, and used today for e.g. speed or rotation measurements, light exposure control, and off-frequency noise filtering. The prototypical chopper uses a rotating wheel with holes that periodically block the incident light beam, with the added feature of being able to control the waveform of the chopped light through the hole diameter to beam width ratio~\cite{Choppers2}.

One may imagine a quantum version of this instrument, with the light beam replaced by a weak coherent state of photons in a one-dimensional channel, and the rotating wheel by a single emitter that periodically couples to the channel. A key difference to the classical optical chopper is of course that a quantum chopper could potentially maintain a unitary evolution of the photons (when disregarding any losses).

As we later discuss, such a quantum chopper could be used 
for single photon pulse shaping~\cite{kolchin}, dynamical routing of single 
photons~\cite{Hoi2011}, and altering of the photon statistics~\cite{Mandel,Bouwmeester}.

Due to the non-linear aspect of the emitter, a quantum chopper may also be able to modulate the statistics of the photons periodically in time. One may even speculate that the resulting periodically modulated signals may be used as input for other quantum optical instruments.

The experimental realization of a quantum chopper seems within the grasp of current nano-photonic technologies that allow for tunable and controllable manipulation of the coupling between different photonic elements. Various tunable coupling schemes have already been proposed and implemented with superconducting qubits, essentially based on the tunability of the Josephson inductance \cite{M-group1,M-group2,M-group3,Delsing}.
Dynamic control has also been demonstrated using an external coupling element between two directly coupled phase and flux qubits \cite{Tun1,Tun2,Tun3,Tun4}, between a phase qubit and a lumped element resonator \cite{Tun5}, and between a 
charge qubit and a coplanar waveguide cavity \cite{Houck}. The latter scheme uses quantum interference to provide an intrinsic method to control the coupling. Recently a qubit architecture that incorporates fast tunable coupling and high coherence has been demonstrated, with dynamical tunability at nanosecond resolution~\cite{M-group5}.

\begin{figure}[tb]
	\centering
	\includegraphics[width=0.8\columnwidth]{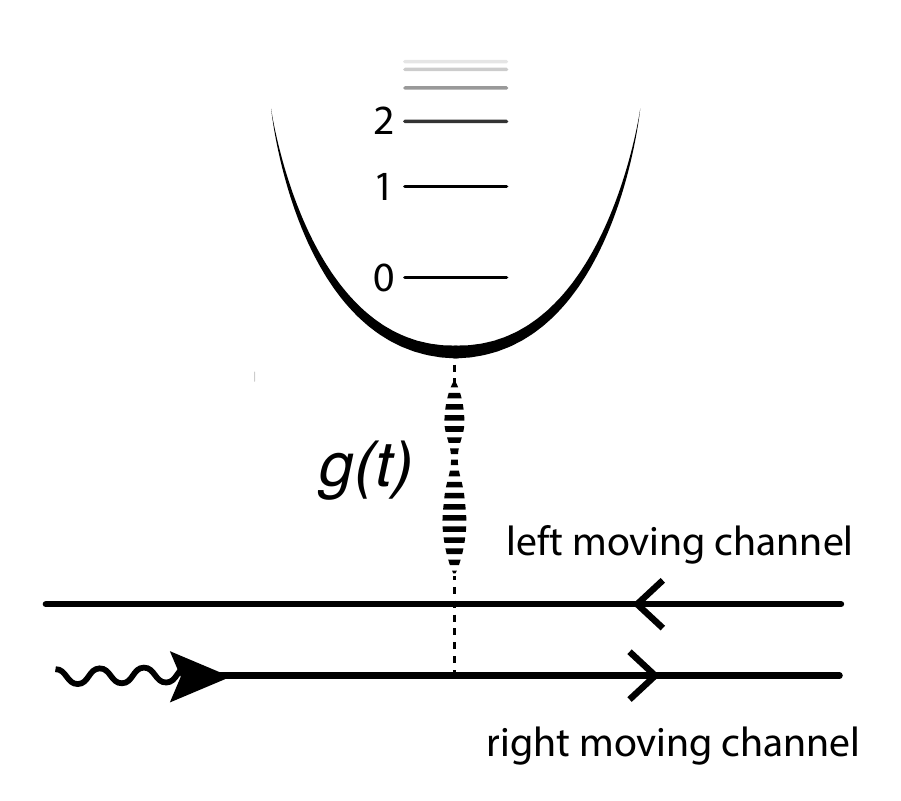}
	\caption{Quantum chopper model consisting of a one-dimensional transmission line supporting two counter-propagating channels, and a cavity with a non-linear spectrum. The two couple through a periodic coupling $g(t)$.}
	\label{fig:figure0}
\end{figure}

We model the proposed quantum analogue of chopper by the following Hamiltonian, 
\begin{align}
H (t)  &= H_0 + V(t) \nonumber \\ &= \int \mathrm{d} \omega \, \hbar \omega (a_\omega^{\dagger} 
a^\pdag_\omega + \tilde{a}_\omega^{\dagger} \tilde{a}^\pdag_\omega) + \hbar 
\omega_c b^{\dagger} b + \frac{U}{2} b^{\dagger 2} b^2 \nonumber \\
&+ \hbar g (t) 
\int \mathrm{d} \omega \, (a_\omega^{\dagger} b + b^{\dagger} a^\pdag_\omega ).
\label{modelHam}
\end{align}
Here $a_{\omega}= (a_{r \omega}+ a_{l \omega})/\sqrt{2}$ and $\tilde{a}_{\omega} = (a_{r \omega}- a_{l \omega})/\sqrt{2}$ describe the two waveguide fields expressed in terms of right- and left-moving modes, $g(t)$ is the coupling strength, and the emitter, described by the bosons $b, b^{\dag}$, has been generalized to a non-linear cavity characterized by a resonance frequency $\omega_c$, and a non-linearity $U$. An illustration of the model is also shown in Figure~\ref{fig:figure0}.

In the next sections we first show how to solve the quasi-stationary dynamics of this system through a generalization of diagrammatic scattering theory to Floquet systems. Then we apply the Floquet scattering theory for describing open Floquet quantum systems explicitly in the few-photon limit. Various results for reflection, transmission and statistics are then summarized. The method is general and can directly be applied to more intricate quantum systems.

\section{Floquet scattering formalism}

An extension of the scattering formalism for time-periodic Hamiltonians was originally proposed in Ref.~\cite{PM} for the calculation of above-threshold-ionization spectra. Remarkably, it offered an effectively time-independent description of the quasistationary limit in terms of the Floquet states. Later, similar scattering approaches have been developed for single-particle scattering~\cite{LR,ER}, many body scattering of non-interacting~\cite{Moskal-b,Moskal1} and interacting~\cite{BC} particles in driven systems.

Let us briefly review the basic ideas of scattering theory. Suppose that at time $t_0 \to - \infty$ we inject $N$ photons into the transmission line, while the cavity is empty. In second quantization, this incoming state is given by ${|p \rangle \equiv | \{ \omega_j \} \rangle | 0 \rangle_c = \left( \prod_{j=1}^N 
a_{\omega_j}^{\dagger} \right) | 0 \rangle | 0 \rangle_c}$, where the vacuum 
state $|0\rangle $ of the transmission line is defined by 
$a_{\omega} | 
0 \rangle = \tilde{a}^\pdag_\omega |0 \rangle = 0$, and 
$|l \rangle_c$ is the photon number state of the cavity, $b^{\dagger} b | l 
\rangle_c = l | l \rangle_c$. The energy of the incoming state equals $\varepsilon_p 
= \sum_{j=1}^N \omega_j$, where we have set $\hbar =1$, as we will continue to do in the rest of this paper. 
After scattering, at time $t \to + \infty$, the cavity is
empty again. Since the Hamiltonian \eqref{modelHam} conserves a number 
of excitations, a scattering state $S | p \rangle$ also contains $N$ photons. Here
$S$ is a scattering operator which emerges from a time 
evolution operator in the long time limit.
In case of the time-independent interaction $V$ 
the energy $\varepsilon_p$ of the input state is conserved in the following sense: 
matrix elements $S_{p'p} = 
\langle p' | S | p \rangle$  appear to be 
proportional to delta functions $\delta (\varepsilon_{p'} - \varepsilon_{p})$, where
$\varepsilon_{p'} =  \sum_{j=1}^N \omega'_j$ is the energy of a state $|  p' \rangle$.
Moreover, $S_{p'p} = \delta_{p'p} - 2 \pi i \delta (\varepsilon_{p'} - 
\varepsilon_{p}) T_{p'p} (\varepsilon_{p})$, where $T_{p'p} (E)$ is the 
energy-dependent $T$ operator containing all the information about scattering off 
the cavity. A systematic way of computing $T (E)$ has been developed in 
\cite{PG} for scatterers with an arbitrary level structure and transition matrix 
elements.

Following the ideas of \cite{PM} we now elaborate on the Floquet scattering 
formalism, particularly adapting it to problems of multi-particle scattering of  (microwave)
photons in one-dimensional waveguides interacting with (artificial) atoms. Our goal is to
present a systematic way of computing the scattering operator $S$ for a 
time-periodic interaction, $V (t) = V (t+T)= \sum_m V^{(m)} e^{-i m \Omega t}$, with 
a fundamental frequency
$\Omega = \frac{2 \pi}{T}$, in the Floquet-extended Hilbert space, thereby generalizing 
the approach of Ref.~\cite{PG} for time-independent couplings. 

We start from an equation for the evolution operator in the interaction picture
\begin{align}
i\frac{d U_{\mathrm{int}} (t, t_0)}{d t} = V_{\mathrm{int}} (t) U_{\mathrm{int}} (t,t_0),
\label{diff_eq}
\end{align}
where $V_{\mathrm{int}} (t) = e^{i H_0 t} V (t) e^{-i H_0 t}$. Taking the limit $t_0 \to -\infty$ we transform \eqref{diff_eq} into the integral form
\begin{align}
 U_{\mathrm{int}}  (t) = \hat{1} - i \int_{-\infty}^t d t' e^{\eta t'} V_{\mathrm{int}} (t') U_{\mathrm{int}} (t'),
\label{int_eq1}
\end{align}
where an infinitesimal factor $\eta > 0$ is additionally introduced for convergence.

Next, we define matrix elements 
$U_{p' p} (t) = \langle p' |  U_{\mathrm{int}}  (t) | p \rangle$ in the eigenbasis $\{ | p \rangle \}$ of $H_0$,  
and express \eqref{int_eq1} in the matrix form
\begin{align}
U_{p' p} (t)  =  \delta_{p' p} &- i \int_{-\infty}^t d t ' \sum_q \sum_m e^{ i (\varepsilon_{p'} - \varepsilon_q - m \Omega - i \eta) t'} 
\nonumber \\
& \times V^{(m)}_{p'q} U_{q p} (t'),
\label{int_eq2}
\end{align}
Being interested in a solution of this equation at times $t>0$, satisfying the condition $\eta t \ll 1$, we look for it in the form
\begin{eqnarray}
U_{p' p} (t) = \delta_{p'p} - \sum_{m'} \frac{e^{i (\varepsilon_{p'} - \varepsilon_p - m' \Omega) t}}{\varepsilon_{p'} - \varepsilon_p - m' \Omega - i \eta} \Theta_{p' p}^{(m')},
\label{ansatz}
\end{eqnarray}
where $\Theta_{p' p}^{(m')}$ are constant matrices. Plugging \eqref{ansatz} into \eqref{int_eq2}, we obtain the equation 
\begin{eqnarray}
\Theta_{p' p}^{(m')}= V_{p' p}^{(m')} -  \sum_q \sum_{n}  \frac{V^{(m'-n)}_{p'q} \Theta_{q p}^{(n)}}{\varepsilon_q - \varepsilon_p - n \Omega - i \eta} ,
\end{eqnarray}
from which we can establish the matrices $\Theta^{(m')}$.

At large times $t$ we make in \eqref{ansatz} the standard replacement $\frac{e^{i \omega t}}{\omega - i \eta} \to 2 \pi i \delta (\omega)$, and thus obtain the scattering matrix
\begin{equation}
S_{p' p} =  \delta_{p'p} - 2 \pi i \sum_{m'} \delta (\varepsilon_{p'} - m' \Omega - \varepsilon_p)  \Theta_{p' p}^{(m')}.
\label{Spp}
\end{equation}
Finally, we introduce the matrix $T_{p'p}^{(m')} (E)$ depending on the energy parameter $E$ and obeying the equation
\begin{align}
T_{p' p}^{(m')} (E)= V_{p' p}^{(m')} +  \sum_q \sum_{n}  \frac{V^{(m'-n)}_{p'q} T_{q p}^{(n)} (E)}{  E - (\varepsilon_q - n \Omega) + i \eta} .
\label{Teq1}
\end{align}
Noticing that  $T_{p'p}^{(m')} (E=\varepsilon_p)$ coincides with the matrix $\Theta_{p' p}^{(m')}$, we arrive at the expression
\begin{equation}
S_{p' p} =  \delta_{p'p} - 2 \pi i \sum_{m'} \delta (\varepsilon_{p'} -m' \Omega - \varepsilon_p)  T_{p' p}^{(m')} (\varepsilon_p) ,
\label{Spp1}
\end{equation}
which relates the $S$ matrix to the $T$ matrix in the time-periodic case. 

As follows from \eqref{Spp1}, the energy $\varepsilon_p$ of an incoming state is conserved 
modulo an integer number of the drive frequency quanta, for each of which we need to find the corresponding $T$ matrix from the equation \eqref{Teq1}. 

Let us consider a generalized version of \eqref{Teq1}
\begin{align}
T_{p'p}^{m'm} (E) = V_{p'p}^{m'm}+ V_{p'q'}^{m'n'} \left[ \frac{1}{E - H'_0 + i \eta} \right]_{q'q}^{n'n} T_{qp}^{nm} (E),
\label{TVT}
\end{align}
where $V_{p'p}^{m'm} \equiv V_{p'p}^{(m'-m)}$, and $H'_0 = H_0 - i \pd_\tau$ is the free Floquet Hamiltonian. The operator $i 
\pd_{\tau}$ is defined by $i \pd_{\tau} |m \rangle = m \Omega |m \rangle$ in terms 
of the Floquet states $|m \rangle = e^{-i m \Omega \tau}$, such that $\langle m' 
| m \rangle =\int_0^{T} \frac{d \tau}{T} e^{i (m'-m) \tau}= \delta_{m'm}$. Thus,  Eq.~\eqref{TVT} is understood
as a relation between operators which act in the Floquet-Hilbert space spanned by $\{|p \rangle \otimes |m \rangle \}$.  
For brevity we 
implicitly assume summations (integrations) over repeated discrete (continuous) 
indices. 

Writing \eqref{TVT} in the operator form $T(E) = V + V (E-H'_0 + i \eta)^{-1} T (E)$, we can easily invert this equation and get
$T (E) =V + V (E-H' + i \eta)^{-1} V$, where $H' = H'_0 + V$ is the full Floquet Hamiltonian. In the matrix representation, this solution reads
\begin{equation}
T_{p'p}^{m'm} (E) = V_{p'p}^{m'm} + V_{p' q'}^{m' n'}
\left[ \frac{1}{E - H' +i \eta} \right]_{q'q}^{n' n} V_{qp}^{nm}.
\label{TVV}
\end{equation}
In turn, the solution of \eqref{Teq1} $T_{p'p}^{(m')} (E) = T_{p'p}^{m'0} (E) $ is obtained from \eqref{TVV} in the special case $m=0$.

Let us make the following important observation: the equation \eqref{TVV} for the $T$ matrix in the time-periodic case has almost the same form as its time-independent counterpart,  the only difference consisting in additional summations over Floquet indices. Noticing that the Hamiltonian \eqref{modelHam} conserves a number of incoming photons after scattering, we can decompose $T = \sum_{N=1}^{\infty} T_N$, where $T_N$ is a normal ordered $N$-photon operator, and straightforwardly generalize the diagrammatic rules of Ref. \cite{PG}. Thus, in the time-periodic case we obtain
\begin{align}
&T^{m' m}_N (E) = \sum_{\{m'_j\}, \{m_j\}} P_{0c} \, \vdots \, V^{m' m^{\phantom{'}}_1} \tilde{G}^{m^{\phantom{'}}_1 m'_1} (E) V^{m'_1 m^{\phantom{'}}_2} \ldots  \nonumber \\
&\times V^{m'_{2N-2}, m^{\phantom{'}}_{2N-1}}  \tilde{G}^{m^{\phantom{'}}_{2N-1},m'_{2N-1}} (E) V^{m'_{2N-1},m } \, \vdots \, P_{0c} ,
\label{TNormalmain}
\end{align}
given by the alternating product of $2N$ interaction operators, $V$, and $2 N-1$ dressed Green's functions, ${\tilde{G} (E) = (E-H'_0 - \Sigma)^{-1}}$, of the cavity. Here ${P_0 }$ is a projector onto the dark (i.e. nonrelaxing) state of the cavity. The Floquet components of the cavity's self-energy $\Sigma^{mm'} \equiv \Sigma^{(m-m')} = - i \pi \sum_n \langle V^{(m-n)} V^{(n-m')} \rangle_{0}$ are given by
an average in the vacuum state of a waveguide. (In particular, for the model \eqref{modelHam} we have $P_{0c} = |0 \rangle_c \,_c\langle 0 | $ and $\Sigma^{mm'}= - i \pi b^{\dagger} b \sum_n g^{(m-n)} g^{(n-m')} $). Finally, the symbol $\vdots (\ldots ) \vdots$ denotes a modified normal ordering, which ignores commutators between field operators contained in different $V$'s, but at the same time obliges to canonically commute a field operator contained in $V$ with $\tilde{G} (E)$ which contains $H_0$.

The expression \eqref{TNormalmain} is exact and sufficient to describe scattering an initial state with arbitrary number of photons. However, because of multiple summations over Floquet indices, it is not optimal for a theoretical analysis. In order to find a more convenient expression, we
transform \eqref{TNormalmain} into the local time representation
\begin{align}
T_{N\varepsilon} (\tau) &\equiv  \sum_{m'} T_N^{(m')} (E) e^{-i m' \Omega \tau} \nonumber \\
&=  \int_0^T \frac{d \tau_1}{T} \ldots \frac{d \tau_{2N}}{T} \delta_T (\tau -\tau_1) \nonumber \\
& \times   P_{0c} \left( \vdots V (\tau_1) \tilde{G}_{\varepsilon} (\tau_1, \tau_2) V (\tau_2) \ldots \right. \nonumber \\
& \left. \ldots V (\tau_{2N-1})  \tilde{G}_{\varepsilon} (\tau_{2N-1},\tau_{2N}) V (\tau_{2N}) \vdots \right) P_{0c} ,
\label{Ttau}
\end{align}
where we introduced the notations $\varepsilon=H_0-E = H_0 - \varepsilon_p$ and
\begin{align}
\tilde{G}_{\varepsilon} (\tau , \tau') = \sum_{m,m'} e^{-i m \Omega \tau} \tilde{G}^{m m'} (E) 
e^{i m' \Omega \tau'},
\label{Gtt}
\end{align}
and used the Poisson resummation formula
\begin{align}
& \sum_{m'} e^{-i m' \Omega (\tau-\tau_1)} \nonumber \\
=& T \sum_{n} \delta (\tau - \tau_1 - n T) \equiv \delta_T (\tau- \tau_1).
\end{align}
Then, from \eqref{Spp1} and \eqref{Ttau} we deduce that the $N$-photon operator contribution to the nontrivial part of the scattering operator equals
\begin{align}
(S-1)_N &=   - i \int_{-\infty}^{\infty} d \tau e^{i (\varepsilon_{p'} - \varepsilon_p) \tau} T_{N \varepsilon} (\tau) \nonumber \\
&=  - i \int_{-\infty}^{\infty} d \tau e^{i H_0 \tau} T_{N \varepsilon} (\tau)  e^{- i  H_0 \tau} ,
\end{align}
and the scattering operator itself is given by
\begin{align}
S = 1 + \sum_{N=1}^{\infty} (- i) \int_{-\infty}^{\infty} d \tau e^{i H_0 \tau} T_{N \varepsilon} (\tau)  e^{- i  H_0 \tau} .
\label{S1T}
\end{align}

Now it is necessary to establish an explicit form of $\tilde{G}_{\varepsilon} (\tau , \tau')$ defined in \eqref{Gtt}. From the relations
\begin{align}
& \sum_{m''} \left[ (m \Omega - \varepsilon) \delta_{m m''} - \Sigma^{m m''} \right] \tilde{G}^{m'' m'} (E) = \delta_{mm'}, \\
& \sum_{m''} \tilde{G}^{m m''} (E) \left[ (m' \Omega - \varepsilon) \delta_{m'' m'} - \Sigma^{m'' m'} \right]  = \delta_{mm'},
\end{align} 
which are equivalent to the definition of $\tilde{G}^{mm'} (E)$, we obtain the differential equations
\begin{align}
& (i \pd_{\tau} - \varepsilon) \tilde{G}_{\varepsilon} (\tau, \tau')- \Sigma (\tau) \tilde{G}_{\varepsilon} (\tau, \tau') = \delta_T (\tau - \tau'), \\
& (-i \pd_{\tau'} - \varepsilon) \tilde{G}_{\varepsilon} (\tau, \tau')-  \tilde{G}_{\varepsilon} (\tau, \tau') \Sigma (\tau') = \delta_T (\tau - \tau'),
\end{align}
where $\Sigma (\tau) = - i \pi \langle V^2 (\tau) \rangle_{0} = \sum_m \Sigma^{(m)} e^{-i m \Omega \tau}$. Equipping them with the periodic boundary conditions in both variables, we find a  solution
\begin{align}
\tilde{G}_{\varepsilon} (\tau, \tau') &= - i T \sum_n \Theta (\tau - \tau' - n T) e^{-i \bar{\varepsilon} (\tau - \tau' - n T)} \nonumber \\
& \times e^{-F_{\mathrm{osc}} (\tau) + F_{\mathrm{osc}} (\tau')},
\end{align}
where $\bar{\varepsilon}  = \varepsilon + \Sigma^{(0)}$ and $F_{\mathrm{osc}} (\tau) = - \sum_{m \neq 0} \frac{\Sigma^{(m)}}{m \Omega} e^{- i m \Omega \tau}$.
Inserting it into \eqref{Ttau} and extending the finite integration ranges $0 < \tau_j <T$ to the infinite ones $-\infty < t_j < \infty$, we cast the scattering operator \eqref{S1T} to the form
\begin{align}
S &= 1 + \sum_{N=1}^{\infty} (-i)^{2N} \int d t_1 \ldots d t_{2N} \Theta (t_1> \ldots > 
t_{2N}) \nonumber \\
&\times e^{i (H_0-E) t_1}  P_{0c} \left( \vvdots \, V 
(t_1) e^{-F (t_1)} e^{F (t_2)} V (t_2) e^{-F (t_2)} \ldots \right. \nonumber \\ 
& \times \left. V (t_{2N-1}) e^{- F (t_{2 N-1})} e^{F (t_{2 N})} V 
(t_{2N}) \, \vvdots \right)  P_{0c},
\label{SNmain}
\end{align}
where $F (t) = i (H_0 +  \Sigma^{(0)} - E) t + 
F_{\mathrm{osc}} (t)$, and $E$ is the energy of an input state. In the following we identify $S$ with $\,_c \langle 0 | S | 0 \rangle_c$.

Note that a $N$-photon operator from the above sum gives only nonzero contribution, if it is applied to a $M$-photon initial state such that $N \leq M$. This means that for a $M$-photon initial state the sum can be truncated after the $M$th term. 

To illustrate an application of \eqref{SNmain} we consider in the next section examples of a single- and two-photon scattering in the model \eqref{modelHam}.

\section{Few photon scattering}
\label{sec:few}

Let us consider the model \eqref{modelHam} and assume that an initial state is prepared in a form of a coherent rectangular pulse of the length $L$, which is initially located far left from the cavity and starts moving towards it in the right direction with a constant velocity $v$. In the interaction picture, this initial state is expressed by
\begin{align}
| \Psi_i \rangle = e^{-|\alpha |^2/2} e^{\alpha \mathcal{A}_{r,\omega_0}^{\dagger}} | 0 \rangle ,
\label{Psi}
\end{align}
where $\mathcal{A}_{r, \omega_0} = \int d \omega \phi (\omega) a_{r \omega} $ is a normalized wavepacket operator centered around the mode $\omega_0$ and broadened over the width $\sim \frac{2 \pi v}{L}$. Formally it is defined by the function
\begin{align}
\phi (\omega ) = \sqrt{\frac{2 v}{\pi L}} \frac{\sin \frac{L}{2 v} (\omega - \omega_0)}{\omega - \omega_0 } , 
\end{align}
which approaches $\sqrt{\frac{2 \pi v}{L}} \delta (\omega - \omega_0)$ for long pulses.

For weak coherence $|\alpha | \ll 1$ we approximate the state \eqref{Psi} by
\begin{align}
| \Psi_i \rangle \approx e^{-|\alpha |^2/2} \left[ 1+ \alpha \mathcal{A}_{r,\omega_0}^{\dagger} + \alpha^2 \frac{(\mathcal{A}_{r,\omega_0}^{\dagger})^2}{2} \right] | 0 \rangle .
\label{Psiapp}
\end{align}
Both single- and two-photon states contributing to \eqref{Psiapp} have a well-defined energy in the long pulse limit $L \to \infty$, and therefore we can apply the scattering operator \eqref{SNmain} to each of them, thus obtaining a final state $| \Psi_f \rangle = S | \Psi_i \rangle$ in the two-photon approximation.

We are interested in computing -- to the leading order in $\alpha$ -- of average transmitted and reflected fields, and their statistical properties quantified by the second order coherence function $g^{(2)}$. In particular, defining the field operators in coordinate representation
\begin{align}
a_{\sigma} (x) = \frac{1}{\sqrt{2 \pi v}} \int d \omega a_{\sigma \omega} e^{i \omega x/v}, \quad \sigma =r,l,
\label{aspace}
\end{align}
we wish to find $\langle \Psi_f | a_{\sigma} (x-v t) | \Psi_f \rangle$ and
\begin{align}
g_{\sigma \sigma'}^{(2)} (t, \tau_d) =& \frac{G_{\sigma \sigma'}^{(2)} (t, \tau_d)}{g_{\sigma}^{(1)} (t) g_{\sigma'}^{(1)} (t+ \tau_d) } , 
\label{g2def}
\end{align}
where
\begin{align}
G_{\sigma \sigma'}^{(2)} (t, \tau_d) =& \langle \Psi_f | a_{\sigma}^{\dagger} (x-v t) a^{\dagger}_{\sigma'} (x-v t- v \tau_d) \nonumber \\
& \times a_{\sigma'} (x-v t - v \tau_d)a_{\sigma} (x-v t)| \Psi_f \rangle, \label{G2def} \\
g_{\sigma}^{(1)} (t) =& \langle \Psi_f | a_{\sigma}^{\dagger} (x-v t) a_{\sigma} (x-v t)| \Psi_f \rangle ,
\label{g1def}
\end{align}
and $\tau_d$ is a delay time. 

Because of an explicit time dependence in the Hamiltonian \eqref{modelHam}, there is no time translational invariance in the long time limit (a corresponding system's state is therefore said to be {\it quasistationary}), and the above defined functions also depend on the evolution time $t$ (though in a periodic way, as we will see later).

Note that the definition \eqref{aspace} implies that the $x$-axis for left-moving photons ($\sigma =l$) points in the left direction.

Since in the Hamiltonian \eqref{modelHam} only even states (${a_{\omega} = \frac{a_{r \omega} + a_{l \omega}}{\sqrt{2}}}$) are coupled to the cavity, and odd states  (${\tilde{a}_{\omega} = \frac{a_{r \omega} - a_{l \omega}}{\sqrt{2}}}$) are decoupled from it,  it appears convenient to express the scattering operator in the basis of even states, also representing the initial state \eqref{Psiapp} in terms of even and odd states. A task of finding $| \Psi_f \rangle$ essentially reduces to evaluation of $S \mathcal{A}_{\omega_0}^{\dagger} | 0 \rangle$ and $S \frac12 (\mathcal{A}_{\omega_0}^{\dagger})^2 | 0 \rangle$, where $\mathcal{A}_{\omega_0}$ is an even counterpart of $\mathcal{A}_{r, \omega_0}$. We consider these cases of single- and two-photon scattering in the following subsections.

\subsection{Single-photon scattering}

Let us first establish how the scattering operator \eqref{SNmain} acts on a single-photon plane wave even state $a^{\dagger}_{\omega} | 0 \rangle$ with energy $E=\omega$. Truncating the sum in \eqref{SNmain} at $N=1$, we obtain
\begin{align}
&  S  a_{\omega}^{\dagger} | 0 \rangle = a_{\omega}^{\dagger} | 0 \rangle \nonumber \\
&- \int d t_1 d t_{2} \Theta (t_1> t_{2}) \int d \omega_1 \int d \omega_2 e^{i (\omega_1-\omega) t_1} \nonumber \\
&\times \,_c \langle 0 | \left( \vvdots \, g 
(t_1)  a^{\dagger}_{\omega_1} b e^{-F (t_1)} e^{F (t_2)} g (t_2)  b^{\dagger} a_{\omega_2} \, \vvdots \right) | 0 \rangle_c \, a_{\omega}^{\dagger} |0 \rangle  \nonumber \\
&= a_{\omega}^{\dagger} | 0 \rangle - \int d \omega_1  \int_{-\infty}^{\infty} d t_1   e^{i (\omega_1-\omega) t_1} g (t_1)    e^{-f_{1\omega} (t_1)} \nonumber \\
& \qquad \times   \int_{-\infty}^{t_1} d t_2 e^{f_{1\omega} (t_2)} g (t_2)  a_{\omega_1}^{\dagger} |0 \rangle  \nonumber \\
&\equiv \int d \omega_1 [\delta_{\omega_1 \omega} + s_{\omega_1 \omega}] a_{\omega_1}^{\dagger} |0 \rangle .
\label{S1main}
\end{align}
The function $F (t)$ for the model \eqref{modelHam} acquires the form
\begin{align}
F (t) = i (H_0 -i \Gamma^{(0)} b^{\dagger} b -E) t + f_{osc} (t) b^{\dagger} b
\end{align}
where $\Gamma^{(0)} + \dot{f}_{osc} (t) = \pi g^2 (t) \equiv \Gamma (t)$, and $f_{osc} (t)$ is fixed by the condition that it does not have a zero frequency component. To single-photon scattering contributes only a single-excitation component $\langle 1 | F (t) | 1\rangle $, and its contribution is appropriately written in terms of the functions $
f_{1 \omega} (t) = i (\omega_c - i \Gamma^{(0)} - \omega) t + f_{osc} (t)$.

Folding \eqref{S1main} with the wavepacket $\phi (\omega)$ and applying the field operator $a (x-v t)$ to the obtained single-photon scattering state, we find
\begin{align}
&  a (x-v t) S \mathcal{A}_{\omega_0}^{\dagger} | 0 \rangle = \frac{e^{-i \omega_0 t_x}}{\sqrt{L} } [1 +2 A (t_x) ] | 0 \rangle ,
\label{aSeven} \\
& A (t_x )= -  \pi g (t_x) e^{-f_1 (t_x)} \int_{-\infty}^{t_x} d t' e^{f_1 (t')} g (t') ,
\label{defA}
\end{align}
where $t$ is time elapsed since the beginning of interaction and $t_x = t -x/v$ is a time lag between the pulse front and the field at point $x$. In \eqref{defA} we have also introduced
\begin{align}
f_1 (t) \equiv f_{1 \omega_0} (t) = - i (\delta + i \Gamma^{(0)}) t + f_{osc} (t),
\label{f1}
\end{align}
with the detuning $\delta = \omega_0 - \omega_c$. 

The function $A (t_x)$ is periodic in its argument, $A (t_x) = A (t_x + T)$, and therefore we can reduce  the central time of pulse evolution $t_x$ (in other words, the observation time at point $x$) to a single period: $t_x \to \tau_c \in [-T/2,T/2]$.

Transforming \eqref{aSeven} to the basis of right and left modes, we obtain the transmitted field (labeled by $r$, the direction of the incident field) and the reflected field (labeled by $l$, the opposite direction)
\begin{align}
&  a_{r,l} (-v t_x) S \mathcal{A}_{r, \omega_0}^{\dagger} | 0 \rangle \nonumber \\
&=  \frac{a (-v t_x)\pm  \tilde{a} (-v t_x)}{\sqrt{2}}  \frac{S \mathcal{A}_{\omega_0}^{\dagger} + \tilde{\mathcal{A}}_{\omega_0}^{\dagger}}{\sqrt{2}} | 0 \rangle \nonumber \\
&= \frac{e^{-i \omega_0 t_x}}{\sqrt{L} }  \left[ \frac{1 \pm 1}{2} + A (t_x) \right] | 0 \rangle.
\end{align}
The transmission $t (\tau_c) = 1+ A (\tau_c)$ and reflection $r (\tau_c) = A (\tau_c)$ amplitudes give envelope shapes of the corresponding fields, and they are not constant in time. Nevertheless, they obey the normalization condition
\begin{align}
\frac{1}{T} \int_0^T d \tau_c \left( | t (\tau_c) |^2 + | r (\tau_c)|^2 \right) =1,
\label{normtr}
\end{align}
corresponding to a conservation of the photon number (see Appendix \ref{NC} for the proof). In the linear regime, one can relate the transmission and reflection amplitudes to the equal-time first order coherences \eqref{g1def} by
\begin{align}
g_r^{(1)} (\tau_c) = \frac{|\alpha|^2}{L} | t (\tau_c) |^2 , \quad g_l^{(1)} (\tau_c) = \frac{|\alpha|^2}{L} | r (\tau_c) |^2 .
\end{align}

Periodic time dependence of an envelope of a scattered field is the main effect of a periodic time modulation of coupling seen in a single-photon scattering. In the following we study this dependence for different modulation protocols. To evaluate $A (\tau_c)$ for $ \tau_c \in [-T/2 , T/2]$ in practice, it is convenient to split the  integral range $[-\infty , \tau_c]$ in \eqref{defA} into two ranges $[-\infty , -T/2]$ and $[ -T/2, \tau_c]$. The integral over the second range can be evaluated numerically, while the integral over the first range can be converted into a geometric series by using the periodicity of $g (t)$ and $f_{osc} (t)$ which results in
\begin{align}
\int_{-\infty}^{-T/2} d t' e^{-i (\delta +i \Gamma^{(0)}) t'} e^{f_{osc} (t')} g(t') = \frac{C_0}{e^{-i (\delta +i \Gamma^{(0)})T} -1}.
\end{align}
Here  $
C_0 =\int_{-T/2}^{T/2} d t' e^{-i (\delta +i \Gamma^{(0)}) t'} e^{f_{osc} (t')} g(t') $ is also evaluated numerically.

Before choosing specific protocols $g (t)$, let us first analyze under which conditions one can expect an interesting time behavior of an envelope $A$.

The most trivial time dependence appears in case of slow driving, when $A (\tau_c)$ instantaneously follows $g (\tau_c)$. It is captured by applying the adiabatic approximation to \eqref{defA}, which is achieved by expanding the integrand close to the upper limit given by the time of observation $\tau_c$. Physically this means that a protocol's history influences very little the present time value of $A$. We have
\begin{align}
A (\tau_c )& = -  \pi g (\tau_c) \int_{-\infty}^{0} d \tau e^{f_1 ( \tau_c + \tau) -f_1 (\tau_c)} g (\tau_c + \tau) \nonumber \\
& \approx - \pi g (\tau_c ) \int_{-\infty}^{0} d \tau e^{\dot{f}_1 ( \tau_c ) \tau } \nonumber \\
& \times [g (\tau_c) + \dot{g} (\tau_c) \tau + \frac12 g (t_x) \ddot{f}_1 ( \tau_c ) \tau^2].
\end{align}
Noticing that $\dot{f}_1 (\tau_c) = - i (\delta + i  \Gamma (\tau_c))$, we conclude
\begin{align}
A (\tau_c ) \approx - \frac{i \Gamma (\tau_c)}{\delta + i \Gamma (\tau_c)} \left[ 1 - \frac{i \dot{g} (\tau_c) }{g (\tau_c)} \frac{\delta - i \Gamma (\tau_c)}{(\delta + i \Gamma (\tau_c))^2} \right].
\label{Ainstan}
\end{align} 
The leading term gives the instantaneous amplitude, and the second term represents the adiabatic correction. This approximation is valid as long as the adiabaticity condition
\begin{align}
\bigg| \frac{ \dot{g} (t) }{g (t)} \bigg|  \ll \sqrt{\delta^2 + \Gamma^2 (t)}
\label{adiab}
\end{align}
is fulfilled. Interesting and unexpected behavior shows up when this condition is violated as we explore in further detail in section~\ref{results}.



\subsection{Two-photon scattering}

Applying the scattering operator \eqref{SNmain} to the two-photon state with energy $E= \omega+ \omega'$ we obtain
\begin{align}
& S a_{\omega}^{\dagger} a_{\omega'}^{\dagger}  | 0 \rangle =  \frac12 a_{\omega}^{\dagger} a_{\omega'}^{\dagger}  | 0 \rangle
+ \int d \omega_1 s_{\omega_1 \omega} a_{\omega_1}^{\dagger} a_{\omega'}^{\dagger}  |0 \rangle \nonumber \\
& +  \int d \omega_1 d \omega_2 d \omega_3 d \omega_4  \int d t_1 d t_2 d t_3 d t_4  \nonumber \\
&\quad  \times \Theta (t_1> t_2 > t_3 > t_{4})  e^{i (H_0-E) t_1}\nonumber \\
&\quad \times   \,_c \langle 0 | \left( \vvdots \, g 
(t_1)  a_{\omega_1}^{\dagger}  b e^{-F (t_1)} e^{F (t_2)} g (t_2) b^{\dagger} a_{\omega_3}  \right. \nonumber \\ 
& \quad \times  e^{-i (H_0 -E) (t_2 - t_3)} g (t_{3}) a_{\omega_2}^{\dagger} b e^{- F (t_{3})} e^{F (t_{4})} g 
(t_4)  b^{\dagger}  a_{\omega_4} \, \vvdots  \nonumber \\
&  \quad +   \vvdots \, g 
(t_1) a^{\dagger}_{\omega_1} b e^{-F (t_1)} e^{F (t_2)} g (t_2) a^{\dagger}_{\omega_2} b e^{-F (t_2)} e^{F (t_3)}  \nonumber \\ 
& \quad \times \left. g (t_{3}) b^{\dagger} a_{\omega_3} e^{- F (t_{3})} e^{F (t_{4})} g 
(t_4) b^{\dagger} a_{\omega_4} \, \vvdots \right)  |0 \rangle_{c} a_{\omega}^{\dagger} a_{\omega'}^{\dagger}  | 0 \rangle 
\nonumber \\
& + (\omega \leftrightarrow \omega').
\label{S2form}
\end{align}
The $N=2$ contribution is represented by the two terms populating the cavity with at most one photon ($\sim bb^{\dagger}bb^{\dagger}$) and with two photons ($\sim bbb^{\dagger}b^{\dagger}$). Simplifying \eqref{S2form} [see Appendix \ref{twophot}] we obtain
\begin{align}
&  a (-v t_x- v \tau_d ) a (-v t_x) S \frac{\mathcal{A}_{\omega_0}^{\dagger \, 2}}{2}  | 0 \rangle =\frac{e^{-i \omega_0 (2 t_x + \tau_d)}}{L} \nonumber \\
&  \times \left[ 1 + 2 A (t_x) + 2 A (t_x + \tau_d) + 4 \bar{B} (t_x , \tau_d) \right] | 0 \rangle,
\label{aaS}
\end{align}
where
\begin{align}
\bar{B} (t_x , \tau_d) &= B (t_x , \tau_d) + A (t_x) A (t_x + \tau_d), \label{bB0} \\
B (t_x , \tau_d)  &= -i U g (t_x) e^{-f_1 (t_x)} g (t_x + \tau_d) e^{-f_1 (t_x + \tau_d)} \nonumber \\
& \times  \int_{-\infty}^{t_x} d t' e^{i U (t'-t_x)+2 f_1 (t')} \frac{A^2 (t')}{g^2 (t')} . \label{B0}
\end{align}

The function $B$ in \eqref{B0} is associated with an inelastic contribution to the two-photon scattering: it vanishes for $U=0$. It is periodic in the argument $t_x$, therefore we can again make a replacement $t_x \to \tau_c$. For the time-independent coupling we recover the expression
\begin{align}
B (\tau_d) = - A^2 \frac{U}{U -2 (\delta + i \Gamma)} e^{i (\delta + i \Gamma) \tau_d} .
\end{align}

In the large $U$ limit, which corresponds to the case of a two-level system, the inelastic contribution \eqref{B0} becomes equal [see Appendix \ref{twophot}]
\begin{align}
B (t_x , \tau_d)  &= -\frac{g (t_x + \tau_d)}{g (t_x)} A^2 (t_x) \nonumber \\
& \times e^{f_{osc} (t_x)-f_{osc} (t_x + \tau_d)} e^{ i (\delta + i \Gamma) \tau_d} .
\label{B0Ui}
\end{align}

With help of \eqref{aaS} we find analogous expressions for transmitted and reflected fields
\begin{align}
&  a_r (-v t_x- v \tau_d ) a_r (-v t_x) S \frac{\mathcal{A}_{r, \omega_0}^{\dagger \, 2}}{2}  | 0 \rangle  \nonumber \\
&  =\frac{e^{-i \omega_0 (2 t_x + \tau)}}{L}  \left[ t (t_x) t (t_x + \tau_d) + B (t_x , \tau_d) \right] | 0 \rangle, \\
&  a_l (-v t_x- v \tau_d ) a_l (-v t_x) S \frac{\mathcal{A}_{r,\omega_0}^{\dagger \, 2}}{2}  | 0 \rangle  \nonumber \\
&  =\frac{e^{-i \omega_0 (2 t_x + \tau_d)}}{L}  \left[ r (t_x) r (t_x + \tau_d) + B (t_x , \tau_d) \right] | 0 \rangle ,
\end{align}
which  allow us to define the corresponding second order coherence functions
\begin{align}
g_{rr}^{(2)} (\tau_c, \tau_d) &= \bigg|1 + \frac{B (\tau_c, \tau_d)}{t (\tau_c) t (\tau_c + \tau_d)} \bigg|^2, \\
g_{ll}^{(2)} (\tau_c, \tau_d) &= \bigg|1 + \frac{B (\tau_c, \tau_d)}{r (\tau_c) r (\tau_c + \tau_d)} \bigg|^2.
\end{align}


\section{Results.}
\label{results}

\subsection{Reflection and transmission}

Assuming a weakly coherent initial signal in the right-moving mode 
$a_{r,\omega_0}$, we study in this section the linear reflection $r (\tau_c)=A (\tau_c)$
and transmission $t (\tau_c) =1 + A (\tau_c)$, which are periodic functions of the 
reduced central time $\tau_c \in [-T/2 , T/2]$. Their absolute values give envelope shapes
of average reflected and transmitted fields, periodically changing in space and time.  This
behavior contrasts with the case time-independent coupling featuring constant 
$r=-\frac{i \Gamma}{\delta + i \Gamma}$ and $t=\frac{\delta}{\delta+ i \Gamma}$.

\begin{figure}[ht]
	\centering
	\includegraphics[width=0.95\columnwidth]{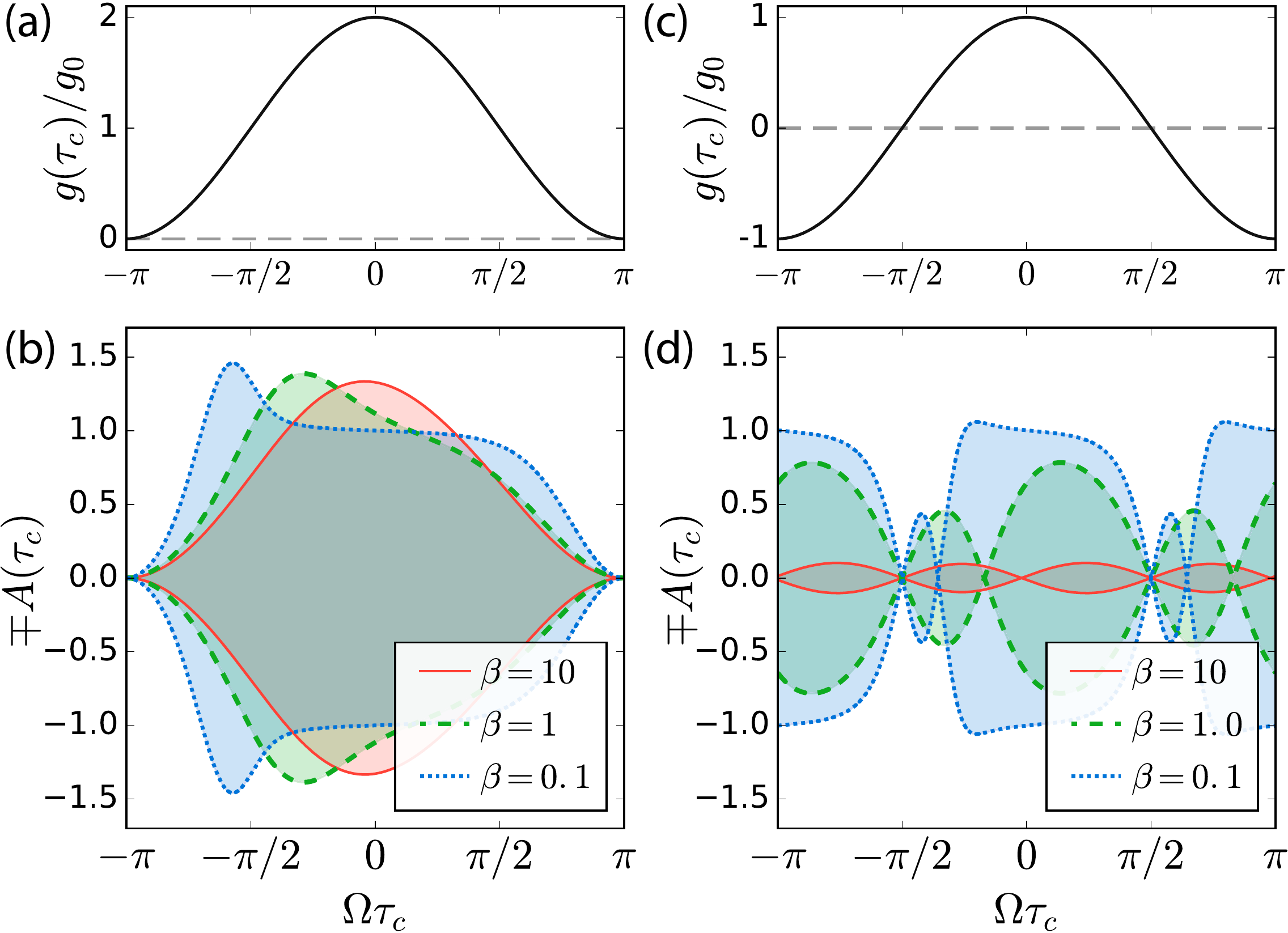}
	\caption{(Color online) The envelopes of the reflected field. 
\textbf{(a)} The ``on-off'' cosine signal, and the resulting \textbf{(b)} 
envelope as a function of the central time $\tau_c$ for various driving speeds. 
Note the perfect transmission ($A=0$) when the coupling is quenched. 
\textbf{(c)} The ``sign change'' cosine signal and the resulting \textbf{(d)} 
envelope. The envelope repeats itself after a half period, and in addition to 
the two coupling quench nodes at $\Omega \tau_c = \pm \pi/2$ an extra node 
develops at $\Omega \tau_c \approx -\pi/2$ (at slow drive) and moves towards 
$\tau_c = 0$ (at fast drive).}
	\label{fig:g1}
\end{figure}

We apply the general results of the Section \ref{sec:few} to two coupling modulation 
protocols:
1) ``on-off'' $g (t) = g_0 (1+\cos \Omega t)$; and 2) ``sign change'' $g (t) = 
g_0 \cos \Omega t$. In the ``on-off'' protocol the coupling strength is 
periodically quenched to zero [Fig.~\ref{fig:g1}(a)], while in the ``sign 
change'' protocol, the sign of $g (t)$ changes after crossing zero 
[Fig.~\ref{fig:g1}(c)]. A notable difference between the two protocols is that 
the former yields a $2 \pi$-periodic modulation of a field's amplitude 
[Fig.~\ref{fig:g1}(b)], while the latter yields a $\pi$-periodic one 
[Fig.~\ref{fig:g1}(d)].

For a time independent interaction, a single photon on resonance ($\delta=0$) 
is fully reflected ($r=-1$), regardless the value of the coupling strength. Should 
the adiabaticity condition \eqref{adiab} be fulfilled at every time $t$ for a time periodic interaction, 
we would 
expect the reflection amplitude $r (t)$ to follow $\Gamma (t)$ instantaneously [see Eq. \eqref{Ainstan}], 
also showing (almost) full reflection in the resonant case (up to a small fraction 
$\sim |\dot{g} (t)/(g (t) \Gamma (t))|$ of the transmitted photon's probability 
density). However, the 
adiabaticity condition \eqref{adiab} is strongly violated for these two protocols.

For any protocol with a momentary quench of coupling this can happen even at slow driving. In these cases the nonadiabatic behavior of $A$ does depend on a protocol's history as we shall see later.

 Moreover, at 
certain time instants the coupling strength in both of them is quenched, 
implying a momentary decoupling of microwave photons from the cavity and hence 
full transmission at these time instants. Since we are dealing with an open 
quantum system, this qualitative picture becomes even more complicated due to 
memory effects, and the non-adiabatic behavior can be explained as a sum over 
histories. Each history has the photon entering the cavity at some initial time, 
$\tau_i$, and leaving at some later time, $\tau_f$, with an amplitude $g(\tau_i) 
g(\tau_f)$, and a weight determined by the decay probability of the photonic 
state in the cavity, $\exp( - \int_{\tau_i}^{\tau_f} \Gamma(\tau) 
\mathrm{d}\tau)$. The reflection coefficient at $\tau_f$, given by the sum over 
initial times $\tau_i$, is highly influenced by the evolution within a memory 
window set by the decay rate of the cavity. 

In the ``on-off'' protocol the memory window is largest for final times after 
the $\Omega \tau_c = -\pi$ node, meaning that the photon remains longer in the 
cavity and is released shortly after when the coupling strength is sufficiently 
increased, producing a spike in the reflection coefficient that overshoots unity 
[Fig.~\ref{fig:g1}(b)].
In the ``sign change'' protocol memory effects create an additional node, that 
is absent in $g (t)$, close to $\Omega \tau_c = -\pi/2$ for slow drive and 
moving towards ${\tau_c = 0}$ for faster drives [Fig.~\ref{fig:g1}(d)]. For 
times shortly after the $-\pi/2$ node of $g (t)$ the memory window includes 
histories with amplitudes of opposite signs, and their competition creates this 
additional node. These two examples show how different protocols may not only chop the 
wavepacket of the incoming photon, but also significantly alter its form.

The resulting envelopes strongly depend on the normalized frequency $\beta = 
\Omega / \Gamma^{(0)}$, where $\Gamma^{(0)}$ is the zeroth harmonic of 
$\Gamma(t)$.
For the fast drive $\beta \gg 1$, we obtain ${A (\tau_c) \approx - \frac23 
(1+\cos \Omega \tau_c)}$ in the ``on-off'' protocol, which means that the 
reflected pulse follows $g (\tau_c)$, not $\Gamma (\tau_c)$;  and ${A (\tau_c) 
\approx - \frac{1}{\beta} \sin 2 \Omega \tau_c}$ following $\Gamma^{(0)} \tau_c 
- f_1 (\tau_c)$ in the ``sign change'' protocol. In the second case, $A 
(\tau_c)$ is negligibly small, so that we have (almost) full transmission 
despite the resonance -- this effect is in sharp contrast to its non-driven 
counterpart, where the full reflection is expected. Thus, this protocol can be 
used for the dynamical routing of photons. For slow drive, $\beta \ll 1$, the adiabaticity condition is 
fulfilled at least within some range around $\tau_c =0$, and this accounts for 
the formation of a plateau with $A (\tau_c) \approx - 1$, resembling full 
reflection in the non-driven resonant case.

As we have seen above, a momentary quench of coupling leads to a formation of nodes in the reflected field. This effect
can be viewed as the quantum version of optical chopping. It is a quantum effect because a scattered single photon remains
in a linear superposition of its transmitted and reflected states. It is analogous to chopping because the amplitude of the transmitted signal is periodically changed from its maximal value down to zero and back again.

\begin{figure}[ht]
	\centering
	\includegraphics[width=\columnwidth]{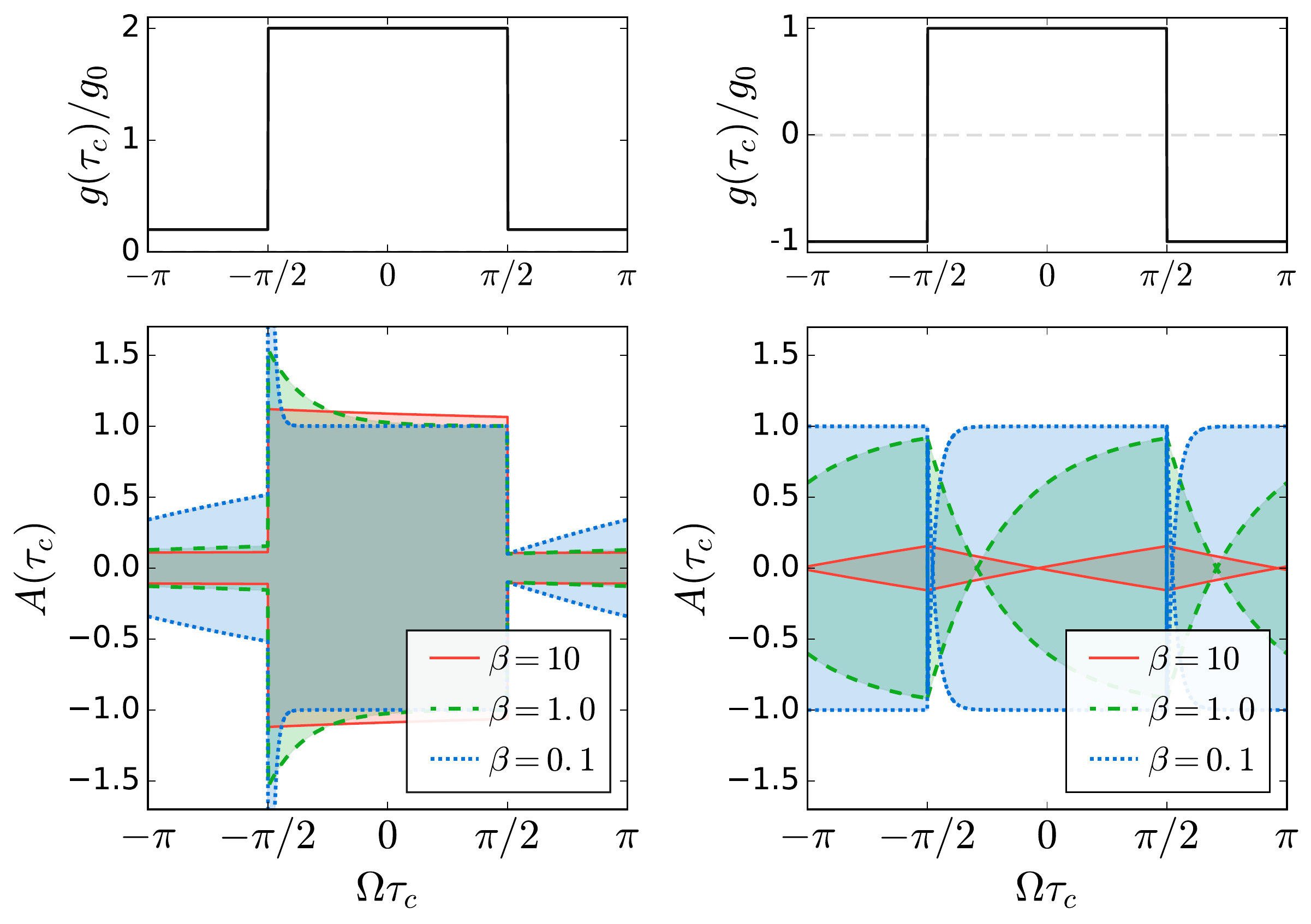}
	\caption{(Color online) Envelope function $A(\tau_c)$ for rectangular driving procedures, which are non-smooth versions of the "on-off" and "sign change" protocols in Fig.~\ref{fig:g1}. The off component of the "on-off" signal has been set at (a more realistic) small non-zero value, $g_{off} = g_0/5$. }
	\label{fig:rect}
\end{figure}

To make this analogy more obvious we show in Fig.~\ref{fig:rect}  the single photon reflection amplitudes in the resonant case $\delta=0$ for rectangular driving procedures that more closely resemble the operation of a conventional chopper:  on the figure at $|g| \sim g_0$ the photon passage is shut (full reflection), while  at $|g| \ll g_0$ it is open (full transmission). The envelope function $A(\tau_c)$ shows qualitatively the same effects as for the cosine signal investigated above. Note that at large $\beta$ (fast drive), the signal shaping also works for the on-off procedure, as it does for the cosine signal.

\subsection{Second order coherence}

The second order coherences \eqref{g2def}  manifest nonlinear effects quantified by the value of $U$. 

Only fast drives, $\beta = \Omega/\Gamma^{(0)} \gg 1$, are able to affect the 
correlations before they decay, and we numerically calculate $g_{ll}^{(2)}$ for 
fast and moderate drives in the two cosine  protocols.

\begin{figure}[b]
\includegraphics[width=0.8\columnwidth]{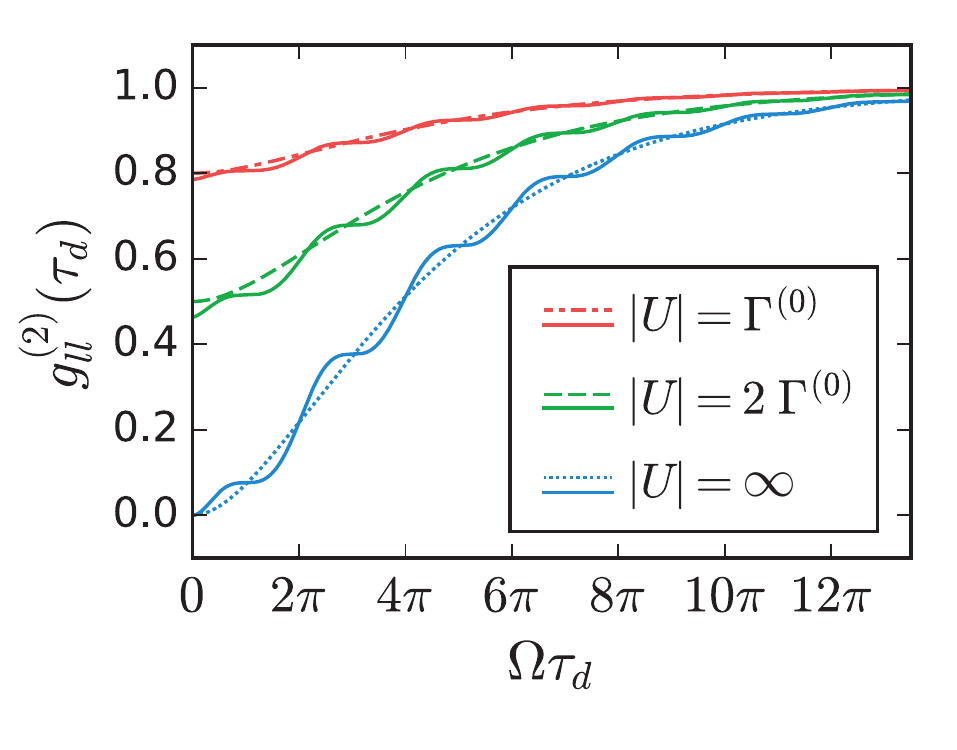}
\caption{(Color online) The $g^{(2)}_{ll}(\tau_c = 0, \tau_d)$ correlation function for the ``on-off'' protocol at fast driving, $\beta=10$. The $g^{(2)}$ correlation for the corresponding non-driven system with a decay rate set to $\Gamma^{(0)}$ are shown as dashed lines, and the (uninteresting) correlations for the driven system slightly oscillate around the non-driven antibunching curves.}
\label{fig:g21}
\end{figure}

In the ``on-off'' protocol, the fast drive only induces small oscillations in 
the correlation function around the non-driven results, as shown in Fig.~\ref{fig:g21}.

\begin{figure}[tb]
	\centering
	\includegraphics[width=\columnwidth]{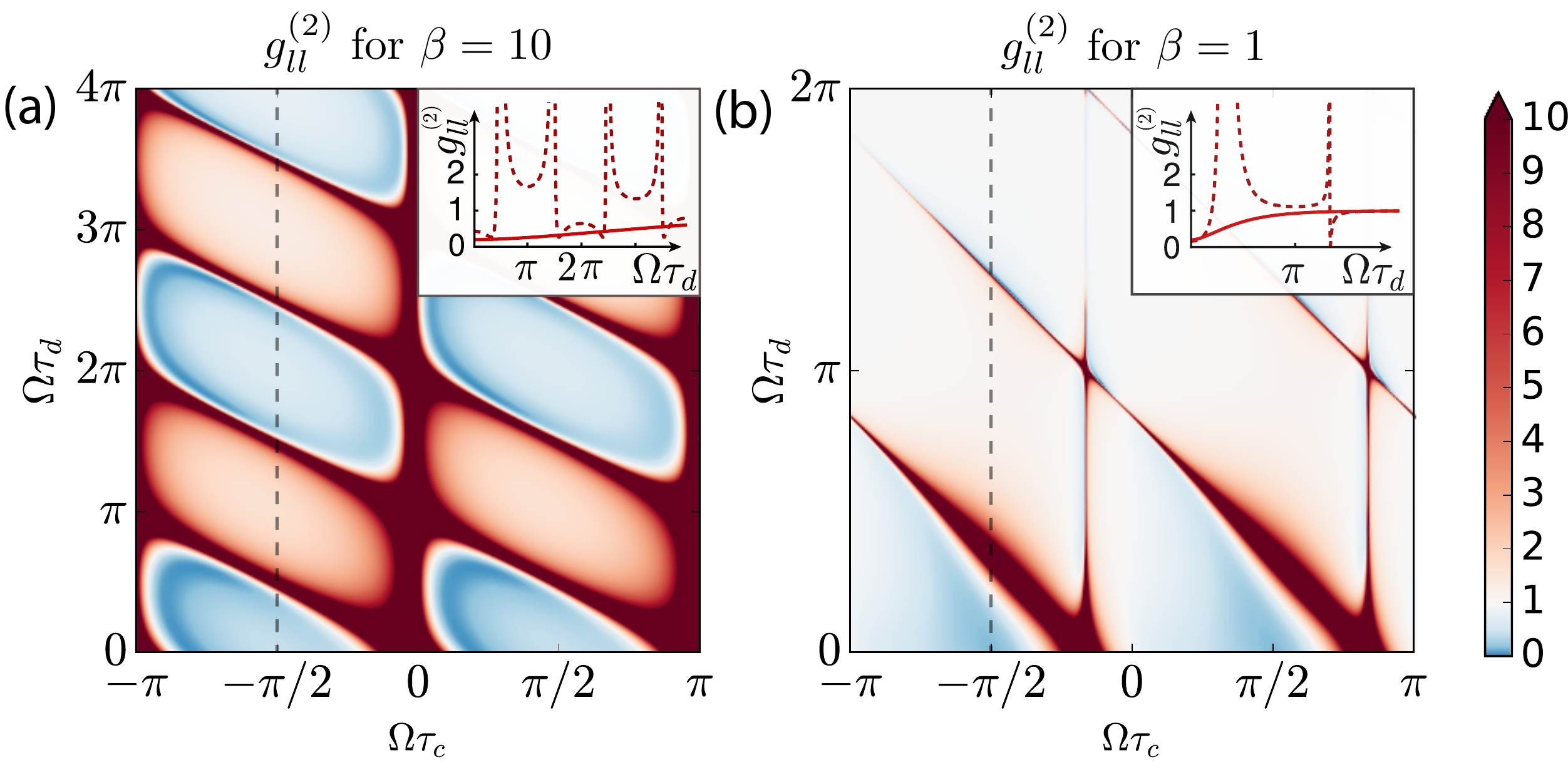}
	\caption{(Color online) Second order coherence of the reflected pulse 
$g_{ll}^{(2)}$ in  the ``sign change'' protocol as a function of central time 
$\tau_c$ and delay time $\tau_d$ for the Kerr  nonlinearity $|U|=4 
\Gamma^{(0)}$. We show results for \textbf{(a)} the fast drive $\beta = 10$, 
where huge periodically repeated bunching peaks are formed and interwoven with 
areas of moderate bunching and anti-bunching; \textbf{(b)} the intermediate 
drive $\beta = 1$, showing the decay of $g_{ll}^{(2)}$ to the uncorrelated value 
(white area). Insets: Comparison of the cuts at $\Omega \tau_c = 
- \frac{\pi}{2}$ (dashed line) with the unmodulated $g_{ll}^{(2)}$ (solid line).}
	\label{fig:g2}
\end{figure}

In 
contrast, the ``sign change'' protocol induces huge bunching effects due to the 
additional node in the single-photon reflection, as can be clearly seen in 
Fig.~\ref{fig:g2}(a). We also find periodic oscillations between strong bunching 
(red areas) and anti-bunching (blue areas) away from $\Omega \tau_c = 0$ and 
$\Omega \tau_c = \pm \pi$. This is a dramatic change in statistical properties 
of the reflected light due to the time dependence of $g (t)$ as compared to the 
case of constant $g$, where $g_{ll}^{(2)}$ is monotonously anti-bunched. For a 
moderate drive, $\beta = 1$, all oscillatory effects in the ``sign change'' 
protocol die out for delay times longer than a single drive period, as shown in 
Fig.~\ref{fig:g2}(b).

\begin{figure}[hb]
	\centering
	\includegraphics[width=0.8\columnwidth]{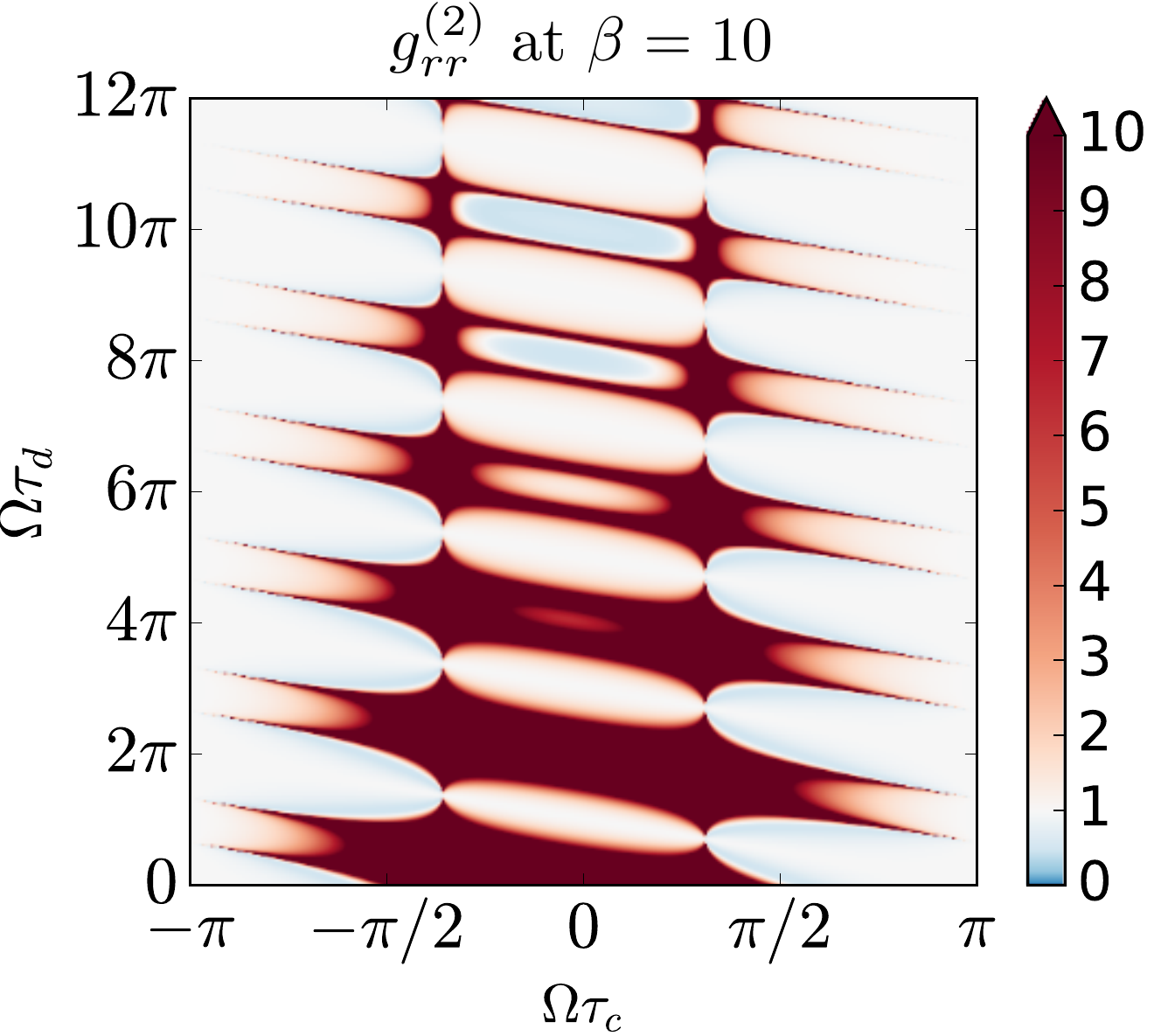}
	\caption{(Color online) The $g^{(2)}_{rr}$ correlation of transmitted light in the ``on-off'' protocol at fast driving, $\beta = 10$, with a nonlinearity $|U|=2\Gamma^{(0)}$. Note the strong periodically recurring bunching due to the wavepacket compression.}
	\label{fig:g23}
\end{figure}

The photon compression by the ``on-off'' driving introduces nodes in the transmission and produces, similarly to the field quench effects in the reflected light for the ``sign change'' protocol, strong bunching in the transmitted light captured by $g^{(2)}_{rr}$. This picture is verified by a numerical calculation of the correlation function for fast drive, $\beta=10$, and nonlinearity, $|U| = 2 \Gamma^{(0)}$ as shown in Fig.~\ref{fig:g23}.

\section{Summary.}

We have proposed a quantum analogue of an optical chopper, operating at the 
few-photon level and realizable by a time-periodic modulation of the 
photon-emitter coupling. We have developed an exact Floquet scattering 
approach based on diagrammatic scattering theory and applied it to 
quantitatively describe scattering of microwave photons from the nonlinear cavity
in two driving protocols of the coupling: ``on-off'' and  ``sign change''. In 
both of them we have observed interesting non-adiabatic memory effects arising 
due to the  driving. In particular, the ``on-off'' protocol produces periodic 
compressions of the photon's wavepacket at slow drive, while at fast drive the signal is directly encoded into the shape of the single photon pulse. The ``sign change'' protocol in turn
gives rise to the additional nodes in the envelope at which the field is 
completely quenched, while at fast drive it may completely change the direction of a photon. These are two examples of chopping realizable at the quantum single-photon level. In addition, in the latter protocol we 
find dramatic changes in statistical properties of the reflected field showing 
up as strong bunching peaks in the $g^{(2)}$ function that are interwoven with 
periodically alternating areas of antibunching and moderate bunching --- 
features that are in sharp contrast to their non-driven counterparts. 
Thus, our findings can be useful  for single photon pulse
shaping, dynamical routing of photons, and
altering of the photon statistics in real time.

\section{Acknowledgements}

We are grateful to A. Fedorov and M. Hafezi for 
useful discussions. The work of V. G. is a part of the Delta-ITP consortium, 
a program of the Netherlands Organization for Scientific Research
(NWO) that is funded by the Dutch Ministry of Education,
Culture and Science (OCW).

\appendix

\section{Proof of the normalization condition}
\label{NC}

To prove the normalization condition \eqref{normtr} we need to show that
\begin{align}
\int_0^T d \tau_c  |A (\tau_c ) |^2 = - \textup{Re} \int_0^T d \tau_c A (\tau_c).
\label{normA}
\end{align}

Let us introduce the function
\begin{align}
W (t) = \int_{-\infty}^t d t' e^{f_1 (t')} g (t') \equiv - \frac{A (t)}{\pi g (t)} e^{f_1 (t)}.
\label{Wfun}
\end{align}
Noticing that $\frac{d}{dt} [f_1 (t) + f_1^* (t) ] =2 \Gamma (t)=2 \pi g^2 (t)$ we integrate lhs of \eqref{normA} by parts
\begin{align}
& \int_0^T d \tau_c \pi \Gamma (\tau_c) e^{-[f_1 (\tau_c) + f_1^* (\tau_c) ]} |W (\tau_c ) |^2 \nonumber \\
=& - \frac{\pi}{2} e^{-[f_1 (\tau_c) + f_1^* (\tau_c) ]} |W (\tau_c ) |^2 \bigg|_0^T  \nonumber \\
&+  \frac{\pi}{2} \int_0^T d \tau_c e^{-[f_1 (\tau_c) + f_1^* (\tau_c) ]} \nonumber \\
& \times [\dot{W}^* (\tau_c ) W (\tau_c ) + W^* (\tau_c ) \dot{W} (\tau_c ) ].
\end{align}
The first term vanishes because of the periodicity of the function $e^{-f_1 (t)} W (t)$, while the second term amounts to
\begin{align}
\frac{\pi}{2} \int_0^T d \tau_c g (\tau_c) [e^{-f_1 (\tau_c)}   W (\tau_c ) + e^{-f_1^* (\tau_c)} W^* (\tau_c ) ],
\end{align}
which coincides with rhs of \eqref{normA}. Thus, \eqref{normtr} is fulfilled.

\section{Evaluation of the two-photon scattering state}
\label{twophot}

The form of the two-photon scattering state \eqref{S2form} can be reduced to
\begin{align}
& S a_{\omega}^{\dagger} a_{\omega'}^{\dagger}  | 0 \rangle =  \frac12 a_{\omega}^{\dagger} a_{\omega'}^{\dagger}  | 0 \rangle
+ \int d \omega_1 s_{\omega_1 \omega} a_{\omega_1}^{\dagger} a_{\omega'}^{\dagger}  |0 \rangle \nonumber \\
& +  \int d \omega_1 d \omega_2   \int d t_1 d t_2 d t_3 d t_4  \nonumber \\
&\quad  \times \Theta (t_1> t_2 > t_3 > t_{4})  e^{i (\omega_1 + \omega_2 -\omega -\omega') t_1}\nonumber \\
&\quad \times    \left(
g (t_1)   e^{-f_{1,E-\omega_2} (t_1)} e^{f_{1,E-\omega_2} (t_2)} g (t_2) \right. \nonumber \\ 
& \quad \times  e^{-i (\omega_2 - \omega' ) (t_2 - t_3)} g (t_{3})  e^{- f_{1 \omega'} (t_{3})} e^{f_{1 \omega'} (t_{4})} g 
(t_4)   \nonumber \\
&  \quad +  2  g 
(t_1)  e^{-f_{1,E-\omega_2} (t_1)} e^{f_{1,E-\omega_2} (t_2)} g (t_2)  e^{-f_2 (t_2)} e^{f_2 (t_3)}  \nonumber \\ 
& \quad \times \left. g (t_{3})  e^{- f_{1 \omega'} (t_{3})} e^{f_{1 \omega'} (t_{4})} g 
(t_4)  \right)   a^{\dagger}_{\omega_1}a^{\dagger}_{\omega_2} | 0 \rangle 
\nonumber \\
& + (\omega \leftrightarrow \omega'),
\end{align}
where $f_2 (t)= i (2 \omega_c + U - 2 i \Gamma^{(0)} -\omega -\omega') t +2 f_{osc} (t)$.
Folding it with $\phi (\omega) \phi (\omega')$ and applying the field operators $a (-v t_x - v \tau_d) a (-v t_x) $ we obtain the expression \eqref{aaS} with
\begin{align}
& 4 \bar{B} (t_x, \tau_d) = \int d \omega_1 d \omega_2   \int d t_1 d t_2 d t_3 d t_4  \nonumber \\
&\quad  \times \Theta (t_1> t_2 > t_3 > t_{4})  e^{i (\omega_1 -\omega_0 ) t_1} e^{ i (\omega_2 - \omega_0) t_2} \nonumber \\
&\quad \times    \left(
g (t_1)   e^{-f_{1} (t_1)} e^{f_{1} (t_2)} g (t_2) \right. \nonumber \\ 
& \quad \times  e^{-i (\omega_2 - \omega_0 ) (t_2 - t_3)} g (t_{3})  e^{- f_{1} (t_{3})} e^{f_{1} (t_{4})} g 
(t_4)   \nonumber \\
&  \quad +  2  e^{-i U (t_2 -t_3)} g 
(t_1)  e^{-f_{1} (t_1)} e^{f_{1} (t_2)} g (t_2)  e^{-2 f_1 (t_2)} e^{2 f_1 (t_3)}  \nonumber \\ 
& \quad \times \left. g (t_{3})  e^{- f_{1} (t_{3})} e^{f_{1} (t_{4})} g (t_4)  \right) \nonumber \\ 
&\quad \times    \left( e^{-i t_x (\omega_1 - \omega_0)} e^{-i (t_x + \tau_d) (\omega_2 - \omega_0)} \right. \nonumber \\
& \quad \left. + e^{-i t_x (\omega_2 - \omega_0)} e^{-i (t_x + \tau_d) (\omega_1 - \omega_0)} \right) ,
\label{S2expl}
\end{align}
and $f_1 (t)$ defined in \eqref{f1}. Performing frequency integrals in \eqref{S2expl} simplifies it to
\begin{align}
& \bar{B} (t_x, \tau_d) =\pi^2   g (t_x + \tau_d) e^{-f_1 (t_x + \tau_d)}  g (t_x)  e^{-f_{1} (t_x)}  \nonumber \\
&\quad \times    \int  d t_2  d t_4  e^{f_{1} (t_{4})} g 
(t_4)  e^{f_{1} (t_2)} g (t_2)    \nonumber \\ 
& \quad \times   [ \Theta (t_x + \tau_d> t_2 > t_x ) \Theta (t_x > t_{4})  \nonumber \\
 & \quad +  2  \Theta (t_x > t_2 > t_4) e^{-i U (t_x -t_2)}     ]  .
 \label{BU1}
\end{align}
The second integral containing the $U$-dependent phase factor can be written in terms of the function \eqref{Wfun} as
\begin{align}
& \int_{-\infty}^{t_x} dt_2 2 \dot{W} (t_2) W (t_2) e^{- i U (t_x -t_2)} \nonumber \\
& = W^2 (t_x) -  i U \int_{-\infty}^{t_x} dt_2 W^2 (t_2) e^{-i U (t_x -t_2)}.
\label{byparts}
\end{align}
Representing
\begin{align}
W^2 (t_x) = \int^{t_x}_{-\infty}   d t_2 e^{f_1 (t_2)} g (t_2) \int^{t_x}_{-\infty}  d t_4  e^{f_1 (t_4)} g (t_4) ,
\end{align}
we substitute \eqref{byparts} in \eqref{BU1} and obtain 
\begin{align}
& \bar{B} (t_x, \tau) =\pi^2   g (t_x + \tau_d) e^{-f_1 (t_x + \tau_d)}  g (t_x)  e^{-f_{1} (t_x)}  \nonumber \\
&\quad \times   \left[  \int^{t_x + \tau_d}_{-\infty} d t_2   e^{f_{1} (t_2)} g (t_2)  \int^{t_x}_{-\infty}  d t_4  e^{f_{1} (t_{4})} g 
(t_4)    \right. \nonumber \\ 
 & \quad \left.  -  i U \int_{-\infty}^{t_x} dt_2 W^2 (t_2) e^{-i U (t_x -t_2)}  \right]  ,
 \label{BU2}
\end{align}
which is equivalent to \eqref{bB0}, \eqref{B0}.

Note that the contribution \eqref{byparts} to the inelastic part of $g^{(2)}$ vanishes in the limit $|U| \to \infty$ (rapid oscillations average the integral in lhs to zero). Thus, we obtain \eqref{B0Ui}.

\end{document}